# Kondo Resonance of a Microwave Photon

Karyn Le Hur[1,2]

[1]*Centre de Physique Théorique, Ecole Polytechnique, CNRS, 91128 Palaiseau Cedex, France*
[2]*Department of Physics, Yale University, New Haven, CT 06520, USA*

We emulate renormalization group models, such as the Spin-Boson Hamiltonian or the anisotropic Kondo model, from a quantum optics perspective by considering a superconducting device. The infra-red confinement involves photon excitations of two tunable transmission lines entangled to an artificial spin-1/2 particle or double-island charge qubit. Focusing on the propagation of microwave light, in the underdamped regime of the Spin-Boson model, we identify a many-body resonance where a photon is absorbed at the renormalized qubit frequency and reemitted forward in an elastic manner. We also show that asymptotic freedom of microwave light is reached by increasing the input signal amplitude at low temperatures which allows the disappearance of the transmission peak.

The asymptotic confinement phenomenon in the infra-red limit is omnipresent in condensed-matter systems and it plays a crucial role in quantum impurity systems, such as the Kondo model describing a single spin-1/2 particle interacting with a bath of conduction electrons [1]. The Kondo effect can also be considered as an example of asymptotic freedom, *i.e.*, the coupling of electrons and spin becomes weak at high temperatures or high energies. This model introduced to describe resistance anomalies in metals with magnetic impurities embodies the "hydrogen atom" of many-body physics [2, 3]. Distinct aspects of this infra-red confinement phenomenon can also be addressed through a one-dimensional boson bath (transmission line) entangling a spin-1/2 particle or two-level system resulting in the Spin-Boson model which can be mapped onto the anisotropic Kondo model and exhibits a plethora of interesting phenomena such as an underdamped-overdamped crossover in the spin dynamics and a quantum phase transition [4–6]. In this Letter, we consider the superconducting Josephson circuit of Fig. 1, which allows to investigate the quantum entanglement in the Spin-Boson model and therefore properties of the anisotropic Kondo model through transport of photons. In the underdamped limit, we prospect to reveal a many-body Kondo resonance in the elastic power of a transmitted microwave photon. This circuit offers the opportunity to export many-body physics in quantum optics.

The superconducting system comprises an artificial spin or double-island charge qubit [7–9] interacting with the zero-point fluctuations of two long one-dimensional transmission lines envisioned from tunable one-dimensional Josephson junction arrays [10–12]. In order to maximize the elastic transmission of a microwave photon, the spin-1/2 object is built from a superconducting double Cooper-pair box where spin up and spin down states refer to the two degenerate charge states $(0,1)$ and $(1,0)$, respectively corresponding to one additional Cooper pair on either island [13]. Recently, geometries involving artificial atoms and transmission lines or cavities have already been realized experimentally [14, 15].

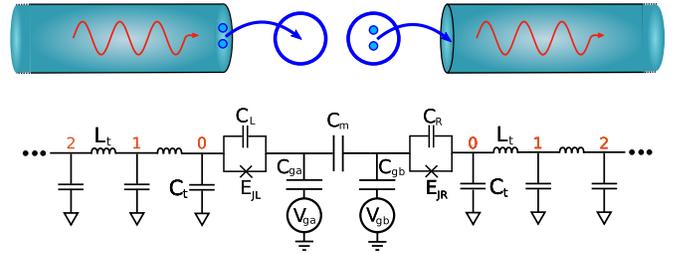

FIG. 1: (Color online) Superconducting circuit envisioned from a double-island charge qubit coupled to two one-dimensional Josephson junction arrays allowing to produce a Kondo resonance in the elastic power of microwave light.

Below, we will assume that the charging energy corresponds to the most dominant term in the Hamiltonian. In fact, close to a charge degeneracy line [7], we can apply the pseudospin representation for the charge states $(0,1)$ and $(1,0)$ reinterpreting them as spin-up and spin-down eigenstates of the operator $\sigma_z$ [13]. The effective detuning $\epsilon = (E_{10} - E_{01}) \to 0$, where $E_{10}$ ($E_{01}$) corresponds to the energy of the spin-down (spin-up) eigenstate, can be adjusted through the gate voltages $V_{ga}$ and $V_{gb}$.

Transfer of Cooper pairs between superconducting islands and leads is described through the Josephson terms $E_{JL}$ and $E_{JR}$ in Fig. 1. In the weak tunneling limit $(E_{JL}, E_{JR}) \ll \min(E_{11} - E_{10}, E_{00} - E_{10})$ one can perform a standard perturbation theory and cotunneling of Cooper pairs then costs an effective energy [13]:

$$E_J = \frac{E_{JL}E_{JR}}{4} \sum_{j=0,1} \left[ \frac{1}{E_{jj} - E_{01}} + \frac{1}{E_{jj} - E_{10}} \right], \quad (1)$$

where $E_{11}$ ($E_{00}$) corresponds to the energy to add (remove) one extra Cooper on the double-island. The Josephson Hamiltonian then takes the form $-(E_J/2)\sigma^+ \exp[i(\phi_l - \phi_r)(x=0)] + h.c.$ [16] where the Josephson phases $\phi_l(x=0)$ and $\phi_r(x=0)$ of the left and

right one-dimensional transmission lines read ($j = l, r$)

$$\phi_j(x=0) = i \sum_{k>0} \frac{2e}{\sqrt{\mathcal{L}c}} \frac{1}{\sqrt{\hbar\omega_k}} (b_{jk} - b_{jk}^\dagger). \quad (2)$$

This Josephson term captures the transmission of a given Cooper pair across the system in Fig. 1. The superconducting reservoirs are explicitly modeled by one-dimensional transmission lines revealing low-energy photon excitations. The left and right transmission lines are described by two distinct sets of harmonic oscillator (photon) operators $b_{lk}$ and $b_{rk}$. Below, we consider the limit where $(C_L, C_R) \ll C_t$ and $(E_{JL}, E_{JR}) \ll \hbar^2/L_T$ in Fig. 1. The spatial solution of the modes can be expressed in terms of the wavevectors $k \approx m\pi/(2\mathcal{L})$, where $m$ is odd for symmetric modes and even for antisymmetric modes, and a transmission line is diagonalized introducing bosonic creation and annihilation operators. Here, $\mathcal{L}$ corresponds to the length of each transmission line and $c = C_t/a$ to the capacitance per unit length; $a$ is the size of a unit cell in each transmission line and we consider the thermodynamic limit $a/\mathcal{L} \to 0$. The photon waves propagate at the speed $v = \omega_c a$ where $\omega_c = 1/\sqrt{L_t C_t}$, the inductances $L_t$ are defined in Fig. 1, and $\omega_k = v|k|$.

To build an explicit analogy with the spin-boson Hamiltonian, we rewrite the Josephson term as a transverse field $H_J = -(E_J/2)\sigma_x$ performing a unitary transformation or spin rotation (see footnote in [29]). Such a procedure, also referred to as a polaron transformation [4, 5], has been applied in the case of a spin-1/2 interacting with the sound modes of a Bose-Einstein condensate [17]. Since the Hamiltonian of the transmission lines does not commute with the spin rotation this produces an effective interaction between the two-level system and the photon excitations. This term can be combined with the capacitive couplings $C_L$ and $C_R$ of Fig. 1. More precisely, since the electrical potential (operator) at the end of a transmission line, i.e., at $x = 0$, takes the form

$$V_j(x=0) = \frac{1}{\sqrt{c\mathcal{L}}} \sum_{k>0} \sqrt{\hbar\omega_k} (b_{jk} + b_{jk}^\dagger), \quad (3)$$

this results in the Spin-Boson Hamiltonian:

$$H = \sum_{j=l,r} \sum_{k>0} \hbar v|k| \left[ b_{jk}^\dagger b_{jk} + \frac{1}{2} \right] - \frac{\epsilon}{2}\sigma_z - \frac{E_J}{2}\sigma_x \quad (4)$$
$$+ \sum_{k>0} \alpha_k \left( -\gamma_l(b_{lk} + b_{lk}^\dagger) + \gamma_r(b_{rk} + b_{rk}^\dagger) \right) \frac{\sigma_z}{2}.$$

The charge operators on the two islands take the forms $Q_b = \frac{2e}{2}(1 + \sigma_z)$ and $Q_a = \frac{2e}{2}(1 - \sigma_z)$. Hereafter the detuning will be fixed to $\epsilon \to 0$ and $\alpha_k = (2e/\sqrt{c\mathcal{L}})\sqrt{\hbar\omega_k}$. The couplings $\gamma_r$ and $\gamma_l$ are given by:

$$\gamma_r = -1 + \frac{C_R}{2} \left( \frac{C_{\Sigma a}}{C_{\Sigma a}C_{\Sigma b} - C_m^2} - \frac{C_m}{C_{\Sigma a}C_{\Sigma b} - C_m^2} \right) \quad (5)$$
$$\gamma_l = -1 + \frac{C_L}{2} \left( \frac{C_{\Sigma b}}{C_{\Sigma a}C_{\Sigma b} - C_m^2} - \frac{C_m}{C_{\Sigma a}C_{\Sigma b} - C_m^2} \right).$$

Following Ref. 7 and the notations of Fig. 1, we have defined the total capacitances seen by each superconducting island: $C_{\Sigma a} = C_L + C_{ga} + C_m$ and $C_{\Sigma b} = C_R + C_{gb} + C_m$.

The analogy with the Spin-Boson model [4, 5] becomes complete when rewriting the Hamiltonian in terms of the symmetric and antisymmetric bosonic combinations:

$$b_{sk} = \cos\theta b_{lk} + \sin\theta b_{rk} \quad (6)$$
$$b_{ak} = \sin\theta b_{lk} - \cos\theta b_{rk}.$$

Choosing $\cos\theta = \gamma_r/\sqrt{\gamma_l^2 + \gamma_r^2}$ and $\sin\theta = \gamma_l/\sqrt{\gamma_l^2 + \gamma_r^2}$, we note that the boson operator $b_{ak}$ only couples to the two-level system through the coupling $\lambda_k = \alpha_k\sqrt{\gamma_l^2 + \gamma_r^2}$. Each transmission line mimics a physical resistor then producing dissipation in the system. In the present circuit, the spectral function of the environment is defined as $J(\omega) = (\pi/\hbar) \sum_{k>0} \lambda_k^2 \delta(\omega - \omega_k) = 2\pi\hbar\alpha\omega e^{-\omega/\omega_c}$ where $\omega_c \gg E_J/\hbar$ represents the high-frequency cutoff of this Ohmic environment [30] and $\alpha$ is given by

$$\alpha = \frac{2R}{R_Q}(\gamma_l^2 + \gamma_r^2). \quad (7)$$

Here, $R_Q = h/(2e)^2$ denotes the quantum of resistance and $R = \sqrt{L_t/C_t}$ is the resistance of each transmission line. It is instructive to observe that in the limit of negligible capacitances $C_L$ and $C_R$ the system naturally converges towards the symmetric condition $\gamma_l = \gamma_r = -1$.

The Spin-Boson Hamiltonian with Ohmic dissipation is intimately related to the Kondo model in the anisotropic regime via bosonization [18]. Other Spin-Boson Hamiltonians such as the Jaynes-Cummings model, in contrast, involve a two-level system interacting with a single mode of a cavity [19]. Other impurity models with photons have also been considered [20, 21]. We are interested in the underdamped regime ($0.1 \preceq \alpha \preceq 0.2$) of the Spin-Boson model where the two-level system displays visible Rabi oscillations but dissipation modifies the qubit frequency which is related to the Kondo energy [4, 5]

$$E_R(\alpha) = \hbar\omega_R = E_J (E_J/\hbar\omega_c)^{\alpha/1-\alpha}. \quad (8)$$

To understand the physical content of the energy $E_R$, it is relevant to apply the unitary transformation $U = \exp(A_l - A_r)$ where $A_j = \sum_{k>0} \frac{\alpha_k \gamma_j}{\hbar\omega_k}(b_{jk}^\dagger - b_{jk})\sigma_z/2$ such that the Hamiltonian can be rewritten as ($\tilde{H} = U^\dagger H U$):

$$\tilde{H} = -\frac{E_J}{2}\sigma^+ e^{i(\Phi_l - \Phi_r)} + h.c. + \sum_{j=l,r} \sum_{k>0} \hbar v|k| \left[ b_{jk}^\dagger b_{jk} + \frac{1}{2} \right], \quad (9)$$

where the phases $\Phi_l = -\gamma_l \phi_l(x = 0)$ and $\Phi_r = -\gamma_r \phi_r(x = 0)$ contain Josephson physics as well as (weak) charging effects. Then, we can define an effective transverse field acting on the dissipative two-level system as $\Delta = E_J \langle \cos(\Phi_l - \Phi_r) \rangle$ such that the artificial atom is described by the effective Hamiltonian $\tilde{H}_{eff} = -(\epsilon/2)\sigma_z - (\Delta/2)\sigma_x$. Bethe ansatz calculations [22, 23] and the adiabatic renormalization [4] in the underdamped limit where $0.1 \preceq \alpha \preceq 0.2$ indeed confirm that $\Delta = E_R$. The bare qubit frequency $E_J/\hbar$ of the two-level system is modified due to the strong renormalization effects associated with the photon bath. One way to experimentally measure the Kondo energy $E_R$ would be through charge measurements since the Fermi liquid ground state imposes that $\langle\sigma_z\rangle \propto \epsilon/E_R$ at small detuning and low temperatures $k_B T \ll E_R$, and the prefactor is accessible from Bethe Ansatz calculations [23].

Below, we show that in the underdamped regime and for temperatures $k_B T \ll E_R$, the Kondo energy $E_R$ can be directly measured based on the (elastic) resonant propagation of a photon. When the system is driven by an external coherent source, the drive, the circuit and the outgoing waves can be treated through the input-output theory [24]. Previous works have studied the limit $\alpha \to 0$ where many-body effects can be fully ignored (the elastic resonance is centered at the bare frequency of the two-level system and converges to a $\delta$-function) [14]. We assume perfect transmission of the microwave signal in the transmission lines such that the input signal reads

$$V_l^{in}(t) = \sum_{k>0} \frac{\alpha_k}{2e} \left( e^{-i\omega_k(t-t_0)} b_{lk}(t_0) + e^{i\omega_k(t-t_0)} b_{lk}^\dagger(t_0) \right). \quad (10)$$

Here, $t_0 < t$ denotes a time in the distance past before any wave packet has reached the two-level system. Similarly, an output field in the left transmission line at time $t_1 > t$ being a time in the distant future after the input field has reached the double-island Cooper box system reads

$$V_l^{out}(t) = \sum_{k<0} \frac{\alpha_k}{2e} \left( e^{-i\omega_k(t-t_1)} b_{lk}(t_1) + e^{i\omega_k(t-t_1)} b_{lk}^\dagger(t_1) \right). \quad (11)$$

Through the Heisenberg relation $\dot{b}_{lk} = (i/\hbar)[H, b_{lk}] = -i\omega_k b_{lk} + (i/2\hbar)\gamma_l \alpha_k \sigma_z$ we relate the properties of the input signal to those of the two-level system.

Below, since we focus on the underdamped limit of the Spin-Boson model which is characterized by a (Rabi) resonance at $\omega = \omega_R$, we establish $\langle\sigma_z(\omega)\rangle \approx \gamma_l \chi(\omega, P_{in}) \langle V_l^{in}(\omega, P_{in})\rangle$ for frequencies in the vicinity of $\omega_R$ where $P_{in} = \langle(V_l^{in})^2\rangle/R$ is the average input power [31]; see EPAPS [25]. Then, we can introduce the reflection coefficient $r(\omega, P_{in}) = \langle V_l^{out}(\omega, P_{in})\rangle / \langle V_l^{in}(\omega, P_{in})\rangle$. Defining the output signal in the right transmission line as

$$V_r^{out} = \sum_{k>0} \frac{\alpha_k}{2e} \left( e^{-i\omega_k(t-t_1)} b_{rk}(t_1) + e^{i\omega_k(t-t_1)} b_{rk}^\dagger(t_1) \right). \quad (12)$$

the transmission coefficient is $t(\omega, P_{in}) = \langle V_r^{out}(\omega, P_{in})\rangle / \langle V_l^{in}(\omega, P_{in})\rangle$. Using the Heisenberg relation of $\dot{b}_{lk}$ with the definitions above we obtain (see

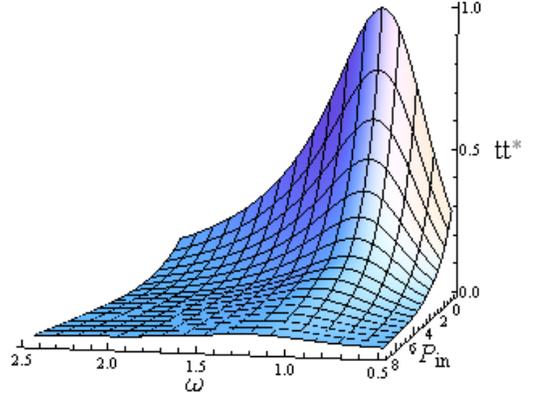

FIG. 2: (Color online) Normalized elastic transmitted power $tt^*(\omega)$ as a function of frequency and driving power for $\gamma_l = \gamma_r$. The parameters are chosen as $\alpha = 0.15$, $\omega_R = 1 = P_R$, $E_J \approx 1.9$, $\omega_c = 50$ and $\hbar = 1$ (we set the ratio $E_J/\hbar\omega_c$ to be moderate, but the "many-body" resonance frequency or renormalized qubit frequency $\omega_R$ is distinct from $E_J/\hbar$).

EPAPS [25]):

$$r(\omega, P_{in}) = \left(1 + \frac{2i\gamma_l^2}{\gamma_l^2 + \gamma_r^2} J(\omega)\chi(\omega, P_{in})\right), \quad (13)$$

$$t(\omega, P_{in}) = -\frac{2i\gamma_r\gamma_l}{\gamma_l^2 + \gamma_r^2} J(\omega)\chi(\omega, P_{in}).$$

First, we consider the linear regime where the amplitude of the input signal is very small, $P_{in} \to 0$ (see footnote in [29]). For frequencies close to the confinement frequency, again assuming the underdamped limit $(0.1 \preceq \alpha \preceq 0.2)$, we derive an expression of the spin susceptibility which agrees with Numerical Renormalization Group results [26]; see EPAPS [25]. This leads to

$$\chi(\omega) = \frac{\omega_R/\hbar}{\omega_R^2 - \omega^2 - i\gamma(\omega)}, \quad (14)$$

where the dissipation factor takes the form $\gamma(\omega) = \omega_R J(\omega)/\hbar$ and is in agreement with the (many-body) Fermi-liquid type ground state [5]. In the linear regime of small input power, we check that the scattering matrix is unitary, $|r|^2 + |t|^2 = 1$, since $J(\omega_R)\Im m\chi(\omega_R) = 1$ showing that the photon propagation is purely elastic close to the resonance (see Fig. 2). We corroborate that the normalized (elastic) power $tt^*(\omega_R)$ flowing to the right transmission line reaches unity since here $\gamma_l \approx \gamma_r$.

In the underdamped regime, the photon propagation across the system is characterized by a many-body resonance at the frequency $\omega_R$ [4–6]: a photon is absorbed at the frequency $\omega_R$ and reemitted forward in a purely elastic manner. In the underdamped regime of the circuit, the qubit is described by a resonance which turns the "photon+Cooper pair" system into an ideal conductor. The phase associated with the reflection coefficient satisfies the following properties. For small $\gamma_l$, the phase vanishes since $V_l^{out} = V_l^{in}$ for an open termination and for $\gamma_r = 0$ the phase is consistent with the Kondo-type $2 \times \pi/2$ phase shift of a right-moving wave.

In fact, the appearance of resonances in such a circuit is not so surprising. For example, let us ignore the Coulomb blockade physics in the two islands completely and replace the Josephson junctions $E_{JL}$ and $E_{JR}$ by purely linear inductances $L_L$ and $L_R$. When $C_L = C_R = C$ and $L_L = L_R = L$, then we corroborate a resonance with $r = 0$ at the frequency $\omega^* = 1/\sqrt{CL + 2C_m L}$; however, we emphasize that $\omega^*$ is distinct from $\omega_R$ which has a many-body origin. In addition, nonlinear effects unavoidably appear in the Josephson circuit of Fig. 1 when increasing the amplitude of the input signal. Under a strong drive, this produces the accumulation of a macroscopic number of photons in the left transmission line which will cause the saturation of the two-level system excitation and the destruction of the resonance peak. More precisely, evaluating the Franck-Condon factor $\langle \cos(\Phi_l - \Phi_r) \rangle = \exp -[\langle(\Phi_l - \Phi_r)^2\rangle/2]$ when increasing the input signal amplitude, we observe that $\langle \Phi_l^2 \rangle$ yields an extra contribution (note, $P_{in} = \langle V_l(x=0)^2\rangle/R$)

$$\frac{1}{(\hbar\omega_R)^2} \sum_{q \in '} \alpha_q^2 \gamma_l^2 \langle b_{lq}^\dagger b_{lq} \rangle = \frac{P_{in} R(2e)^2 \gamma_l^2}{(\hbar\omega_R)^2}, \quad (15)$$

and the symbol ' refers to momenta such that $\omega_q \sim \omega_R$ in the case of a monochromatic signal with a frequency close to $\omega_R$. The Josephson process in Eq. (9) becomes exponentially diminished and this results in an exponential suppression of $\Im m \chi(\omega = \omega_R)$ (see EPAPS [25])

$$\Im m \chi(\omega_R, P_{in}) J(\omega_R) = \exp - \left( \frac{P_{in}}{P_R} \frac{R}{R_Q} \pi \gamma_l^2 \right), \quad (16)$$

where $P_R = \hbar \omega_R^2$. The nonlinearity of the two-level system produces an exponential decrease of the spin susceptibility at $\omega = \omega_R$. Then, this causes the disappearance of the Rayleigh transmission resonance; see Fig. 2. When $P_{in} \gg P_R$ one reaches the asymptotic freedom of microwave light where $r(\omega_R) \sim 1$; see Eqs. (13) and (16). These nonlinear effects are driven by the Josephson-type Hamiltonian in Eq. (9). Increasing the driving power $P_{in}$, the scattering matrix becomes non-unitary since $J(\omega_R) \Im m \chi(\omega_R, P_{in}) < 1$ (which hides the presence of additional inelastic Raman corrections [14]). An open question concerns the inelastic Raman spectrum, which should become prominent in the overdamped limit of the Spin-Boson model. The asymptotic freedom of light, resulting in $r(\omega_R) \sim 1$ can also be reached when increasing the temperature producing a prominent decoherence of the two-level system and a strong decrease of $\langle \sigma_x \rangle$ and of the Josephson coupling in Eq. (9); see EPAPS [25].

To summarize, in the underdamped regime of the Spin-Boson model, sending a microwave photon produces a many-body Kondo resonance. For moderate and accessible values of the dissipation strength $0.1 \preceq \alpha \preceq 0.2$, the confinement frequency $\omega_R$ is clearly distinguishable from the bare Josephson frequency $E_J/\hbar$ of the two-level atom as a result of the continuum of photon modes in the (very long) transmission lines. The width of the resonance peak reflects the Fermi-liquid type ground state. We assumed that the detuning $\epsilon$ and thermal effects through $k_B T$ are smaller than the Kondo energy $E_R$, and in the underdamped regime, $E_R$ is not too small compared to the Josephson energy $E_J$. For Josephson junction arrays with large resistances, this circuit would offer the opportunity to study overdamped spin dynamics and quantum phase transitions using nonlinear optics. Note that the development of techniques such as the numerical renormalization group [27] or a stochastic-type approach [28] would be necessary to combine the traditional input-output theory with many-body physics of quantum impurity models. Finally, such superconducting quantum devices can be used for controllable (single-)photon sources in which a plethora of novel effects related to many-body physics and nonlinear quantum optics can be realized.

This work was supported by DOE, under the grant DE-FG02-08ER46541. We thank M. Devoret for discussions.

# Supplementary Material: Kondo Resonance of a Microwave Photon

Karyn Le Hur[1,2]

[1]*Centre de Physique Théorique, Ecole Polytechnique, CNRS, 91128 Palaiseau Cedex, France*
[2]*Department of Physics, Yale University, New Haven, CT 06520, USA*

We introduce the input-output theory and we derive an analytical expression of the dynamical spin susceptibility in the weak-coupling limit in the presence of an incoming signal which illustrates the concepts of the Renormalization Group and agrees with the Numerical Renormalization Group. We also investigate nonlinear effects close to the Kondo (resonance) frequency or renormalized qubit frequency $\omega_R$ when increasing the amplitude of the input signal.

PACS numbers: 03.65.Yz, 03.75.Lm, 42.50.-p, 85.25.-j

## INPUT-OUTPUT ANALYSIS

First, we apply the concepts issued from input-output theory for the left transmission line and we derive Eqs. (13) in the main text. Let us start with the Heisenberg equations of motion:

$$\begin{aligned}\dot{b}_{lk} &= \frac{i}{\hbar}[H, b_{lk}] \\ &= -i\omega_k b_{lk} + \frac{i}{2\hbar}\gamma_l \alpha_k \sigma_z(t).\end{aligned} \quad (1)$$

Note that the original Hamiltonian and H defined in the main text are related through a unitary transformation analogous to $U$. Now, $\dot{b}_{lk} + \dot{b}_{lk}^\dagger = (i/\hbar)[UHU^{-1}, b_{lk} + b_{lk}^\dagger] = (i/\hbar)[H, b_{lk} + b_{lk}^\dagger]$ with $\dot{b}_{lk} = db_{lk}/dt$ where $U$ has been defined in the main text. Therefore, $\dot{b}_{lk} = (i/\hbar)[H, b_{lk}]$.

By integrating the Heisenberg equations of motion, this results in:

$$\begin{aligned}b_{lk}(t) &= \exp[-i\omega_k(t-t_0)]b_{lk}(t_0) + \frac{i}{2\hbar}\gamma_l\alpha_k \int_{t_0\to-\infty}^t d\tau e^{-i\omega_k(t-\tau)}\sigma_z(\tau) \\ b_{lk}^\dagger(t) &= \exp[i\omega_k(t-t_0)]b_{lk}^\dagger(t_0) - \frac{i}{2\hbar}\gamma_l\alpha_k \int_{t_0}^t d\tau e^{i\omega_k(t-\tau)}\sigma_z(\tau).\end{aligned} \quad (2)$$

$\omega_k$ is the bare frequency of the transmission line (when setting $\gamma_l = 0$), and therefore we identify $t_0 < t$ as a time in the distance past before any wave packet has reached the qubit (two-level system). Assuming perfect transmission of the signal in the left transmission line then we identify:

$$V_l^{in}(t) = V_l(x=0,t) = \sum_k \frac{\alpha_k}{2e}\left(\exp[-i\omega_k(t-t_0)]b_{lk}(t_0) + \exp[i\omega_k(t-t_0)]b_{lk}^\dagger(t_0)\right). \quad (3)$$

Now, let us consider an output field in the left transmission line at time $t_1 > t$ being a time in the distant future after the input field has reached the qubit. In this case:

$$\begin{aligned}b_{lk}(t) &= \exp{-i\omega_k(t-t_1)}b_{lk}(t_1) - \int_t^{t_1\to+\infty} d\tau \frac{i}{2\hbar}\gamma_l\alpha_k e^{-i\omega_k(t-\tau)}\sigma_z(\tau) \\ b_{lk}^\dagger(t) &= \exp{i\omega_k(t-t_1)}b_{lk}^\dagger(t_1) + \int_t^{t_1\to+\infty} d\tau \frac{i}{2\hbar}\gamma_l\alpha_k e^{i\omega_k(t-\tau)}\sigma_z(\tau).\end{aligned} \quad (4)$$

Then, identifying:

$$V_l^{out}(t) = V_r(x=0,t) = \sum_k \frac{\alpha_k}{2e}\left(\exp[-i\omega_k(t-t_1)]b_{lk}(t_1) + \exp[i\omega_k(t-t_1)]b_{lk}^\dagger(t_1)\right), \quad (5)$$

after Fourier transformation and comparing Eqs. (2) and (4), we get:

$$\begin{aligned}V_l^{out}(\omega) &= V_l^{in}(\omega) + i\frac{\gamma_l}{\gamma_l^2+\gamma_r^2}J(\omega)\sigma_z(\omega) - i\frac{\gamma_l}{\gamma_l^2+\gamma_r^2}J(\omega)\sigma_z^*(\omega) \\ (V_l^{out})^*(\omega) &= (V_l^{in})^*(\omega) - i\frac{\gamma_l}{\gamma_l^2+\gamma_r^2}J(\omega)\sigma_z^*(\omega) + i\frac{\gamma_l}{\gamma_l^2+\gamma_r^2}J(\omega)\sigma_z(\omega).\end{aligned} \quad (6)$$

(We check that the results are unchanged restricting the sum in the expression of $V_l^{in}$ to $\sum_{q>0}$ and the sum in the expression of $V_l^{out}$ to $\sum_{q<0}$. We can always change $q \to -q > 0$ in the sum for $V_l^{out}$.)

Defining $\langle \sigma_z(\omega)\rangle = \chi(\omega, P_{in})\gamma_l \langle V_l^{in}(\omega, P_{in})\rangle$ and combining the two lines then we find

$$\langle V_l^{out}(\omega, P_{in})\rangle = \langle V_l^{in}(\omega, P_{in})\rangle \left(1 + 2i\frac{\gamma_l^2}{\gamma_l^2 + \gamma_r^2} J(\omega)\chi(\omega, P_{in})\right). \tag{7}$$

The reflection coefficient in the main text then is defined as:

$$r(\omega, P_{in}) = \frac{\langle V_l^{out}\rangle}{\langle V_l^{in}\rangle} = 1 + 2i\frac{\gamma_l^2}{\gamma_l^2 + \gamma_r^2} J(\omega)\chi(\omega, P_{in}). \tag{8}$$

Note also that $\sigma_z^*(\tau) = \sigma_z(\tau)$ implies $\chi^*(\omega) = \chi(-\omega)$.

Proceeding in the same way with $\dot{b}_{rk}$ then we obtain:

$$t(\omega, P_{in}) = -\frac{2i\gamma_r\gamma_l}{\gamma_l^2 + \gamma_r^2} J(\omega)\chi(\omega, P_{in}). \tag{9}$$

### DYNAMICAL SUSCEPTIBILITY IN UNDERDAMPED AND LINEAR REGIME

Here, we derive an expression of the dynamical spin susceptibility $\chi(\omega)$ for frequencies $\omega$ close to the confinement frequency $\omega_R$ based on a weak-coupling expansion assuming a negligible input power. We apply a procedure analogous to the Non-Interacting Blip Approximation which has been shown to be valid in the limit $0.1 \preceq \alpha \preceq 0.2$ [1–3].

Let us start with the Heisenberg equation of motion where the Hamiltonian $H$ is given in Eq. (4) of the main text and evaluate $\langle \sigma_z(\omega)\rangle$ ($H$ and the original Hamiltonian are related through unitary transformation which commutes with the operator $\sigma_z$):

$$\dot{\sigma}_z = \frac{i}{\hbar}[H, \sigma_z] = -\frac{E_J}{\hbar}\sigma_y. \tag{10}$$

Then, we obtain:

$$\dot{\sigma}_y = \frac{E_J}{\hbar}\sigma_z - \left(\frac{\gamma_l}{\hbar}\sum_{k>0}\alpha_k(b_{lk} + b_{lk}^\dagger) - \frac{\gamma_r}{\hbar}\sum_{k>0}\alpha_k(b_{rk} + b_{rk}^\dagger)\right)\sigma_x. \tag{11}$$

For simplicity, here we assume that $\epsilon = 0$ strictly. From these two equations, we obtain:

$$\ddot{\sigma}_z = -\frac{E_J}{\hbar}\dot{\sigma}_y = -\frac{E_J}{\hbar}\left(\frac{E_J}{\hbar}\sigma_z - \left(\frac{\gamma_l}{\hbar}\sum_{k>0}\alpha_k(b_{lk} + b_{lk}^\dagger) - \frac{\gamma_r}{\hbar}\sum_{k>0}\alpha_k(b_{rk} + b_{rk}^\dagger)\right)\sigma_x\right). \tag{12}$$

As a result:

$$\ddot{\sigma}_z + \left(\frac{E_J}{\hbar}\right)^2 \sigma_z = +\frac{E_J}{\hbar}\left(\frac{\gamma_l}{\hbar}\sum_{k>0}\alpha_k(b_{lk} + b_{lk}^\dagger) - \frac{\gamma_r}{\hbar}\sum_{k>0}\alpha_k(b_{rk} + b_{rk}^\dagger)\right)\sigma_x. \tag{13}$$

On the other hand, we get:

$$\dot{\sigma}_x = +\left(\frac{\gamma_l}{\hbar}\sum_{k>0}\alpha_k(b_{lk} + b_{lk}^\dagger) - \frac{\gamma_r}{\hbar}\sum_{k>0}\alpha_k(b_{rk} + b_{rk}^\dagger)\right)\sigma_y. \tag{14}$$

This is equivalent to:

$$\sigma_x(t) = \sigma_x(t_i) - \frac{\hbar}{E_J}\int_{t_i}^{t} dt' \left(\frac{\gamma_l}{\hbar}\sum_{k>0}\alpha_k(b_{lk} + b_{lk}^\dagger)(t') - \frac{\gamma_r}{\hbar}\sum_{k>0}\alpha_k(b_{rk} + b_{rk}^\dagger)(t')\right)\dot{\sigma}_z(t'). \tag{15}$$

During the integration procedure, $t_i < t$ represents an initial time which is set arbitrarily (but at $t = t_i$ we must satisfy that the incoming signal has already reached the two-level system since we want to study the spin response to an incoming microwave signal). This results in:

$$\ddot{\sigma}_z(t) + \left(\frac{E_J}{\hbar}\right)^2 \sigma_z(t) = \frac{E_J}{\hbar}\left(\frac{\gamma_l}{\hbar}\sum_{k>0}\alpha_k(b_{lk}+b_{lk}^{\dagger})(t) - \frac{\gamma_r}{\hbar}\sum_{k>0}\alpha_k(b_{rk}+b_{rk}^{\dagger})(t)\right)\sigma_x(t_i) \quad (16)$$

$$- \left(\frac{\gamma_l}{\hbar}\sum_{k>0}\alpha_k(b_{lk}+b_{lk}^{\dagger})(t) - \frac{\gamma_r}{\hbar}\sum_{k>0}\alpha_k(b_{rk}+b_{rk}^{\dagger})(t)\right)$$

$$\times \int_{t_i}^{t} dt'\left(\frac{\gamma_l}{\hbar}\sum_{q>0}\alpha_q(b_{lq}+b_{lq}^{\dagger})(t') - \frac{\gamma_r}{\hbar}\sum_{q>0}\alpha_q(b_{rq}+b_{rq}^{\dagger})(t')\right)\dot{\sigma}_z(t').$$

If $V_l^{in} \to 0$, at the time $t_i$, we can assume that the spin-boson model is in its ground state and therefore from Bethe Ansatz we rigorously identify [4] $\langle\sigma_x(t_i)\rangle = \langle\sigma_x\rangle \to E_R/E_J$ (again, assuming that we are in the weak-coupling regime with $0 \le \alpha \ll 1/3$; for another derivation, see below). More precisely, we obtain the exact equation (for all $\alpha$) [4, 5]:

$$\langle\sigma_x\rangle = \frac{1}{2\alpha-1}\frac{E_J}{\hbar\omega_c} + \mathcal{C}(\alpha)\frac{E_R}{E_J}, \quad (17)$$

where

$$\mathcal{C}(\alpha) = \frac{e^{-b/(2-2\alpha)}\Gamma[1-1/(2-2\alpha)]}{\sqrt{\pi}(1-\alpha)\Gamma[1-\alpha/(2-2\alpha)]}, \quad (18)$$

$\Gamma$ is the incomplete gamma function and $b = \alpha\ln\alpha + (1-\alpha)\ln(1-\alpha)$. Then, we apply the mean-field decoupling:

$$\frac{E_J}{\hbar}\left\langle\left(\frac{\gamma_l}{\hbar}\sum_{k>0}\alpha_k(b_{lk}+b_{lk}^{\dagger})(t) - \frac{\gamma_r}{\hbar}\sum_{k>0}\alpha_k(b_{rk}+b_{rk}^{\dagger})(t)\right)\sigma_x(t_i)\right\rangle \approx \frac{E_R}{\hbar}\frac{\gamma_l}{\hbar}V_l^{in}(t). \quad (19)$$

The mean-field decoupling (averaging) is usually valid in the underdamped limit of the spin-boson model [3], and then the input signal $V_l^{in}(t) = V_0\cos(\omega t)$ then (simply) mimics a time-dependent detuning $\epsilon(t)$ acting on the two-level system. This approximation is well justified for time scales smaller than the time $\sim 1/\omega_R$ which corresponds to the crossover scale toward the strong-coupling regime. Here, we extend this approximation for frequencies slightly below the confinement frequency. This will result in $\sigma_z(\omega) \approx \gamma_l\chi(\omega)V_l^{in}(\omega)$ in the vicinity of the confinement frequency.

Now, let us focus on the (real part of the) last term in Eq. (16). Assuming $t_i - t \to -\infty$, this gives:

$$-\frac{\gamma_l^2+\gamma_r^2}{\hbar^2}\int_{-\infty}^{t}dt'\sum_{k>0}\left(\alpha_k^2\langle b_{lk}b_{lk}^{\dagger}\rangle e^{-i\omega_k(t-t')} + \alpha_k^2\langle b_{lk}^{\dagger}b_{lk}\rangle e^{i\omega_k(t-t')}\right)\langle\dot{\sigma}_z(t')\rangle \quad (20)$$

$$\to -\frac{\gamma_l^2+\gamma_r^2}{\hbar^2}\int_{-\infty}^{t}dt'\sum_{k>0}\alpha_k^2\coth\left(\frac{\beta\hbar\omega_k}{2}\right)\cos(\omega_k(t-t'))\langle\dot{\sigma}_z(t')\rangle,$$

where $\beta = 1/k_BT$. When $P_{in} = \langle(V_l^{in})^2\rangle/R = V_0^2/2R \to 0$, the boson modes are taken to be in thermal equilibrium. After Fourier transformation (assuming $t \sim 0$), then to compute the real part one needs to evaluate:

$$-\frac{1}{\hbar\pi}\omega^2\mathcal{P}\left(\int_0^{\omega_c\to+\infty}d\omega'\frac{1}{\omega'^2-\omega^2}J(\omega')\coth\left(\frac{\beta\hbar\omega'}{2}\right)\right)\langle\sigma_z(\omega)\rangle, \quad (21)$$

and $\mathcal{P}$ denotes the principal part of the integral.

In particular, for frequencies $\omega \sim E_J/\hbar$ which corresponds to the resonance (Rabi) frequency of the non-dissipative spin and for $T \to 0$, this renormalizes the term $(E_J/\hbar)^2\langle\sigma_z(\omega)\rangle$ in $\omega_R^2\langle\sigma_z(\omega)\rangle$ where $\omega_R = E_R/\hbar$ formally obeys:

$$\omega_R^2 = \frac{E_R^2}{\hbar^2} = \left(\frac{E_J}{\hbar}\right)^2 - 2\alpha\left(\frac{E_J}{\hbar}\right)^2\ln\left(\frac{\hbar\omega_c}{E_J}\right). \quad (22)$$

The present approach assumes a weak-coupling $\alpha$ for which $\omega_R$ can be identified to the Kondo frequency:

$$\omega_R \approx \frac{E_J}{\hbar}\left(\frac{E_J}{\hbar\omega_c}\right)^{\frac{\alpha}{1-\alpha}}. \quad (23)$$

This formula is formally valid for $k_B T \ll E_J$ since the dominant contribution stems from frequencies higher than $E_J/\hbar$. The renormalization procedure in the Spin-Boson model, consisting to integrate out high-frequency photon modes, results in an effective Josephson energy $E_R < E_J$ [1].

To compute the imaginary part leading to dissipation (damping), first let us consider that $k_B T > E_R$, such that the weak-coupling decoupling in Eq. (19) becomes rigorous. For frequencies $\omega \sim \omega_R$, then we find a damping term

$$-i\gamma(\omega, T)\langle\sigma_z(\omega)\rangle, \tag{24}$$

where

$$\gamma(\omega, T) = \frac{k_B T}{\hbar^2} J(\omega). \tag{25}$$

We check that $\gamma$ is odd in frequency which guarantees that $\chi^*(\omega) = \chi(-\omega)$. Now, let us decrease progressively the temperature such that $k_B T \sim E_R$. Then, the damping term takes the quantum form:

$$\gamma = \frac{\omega_R}{\hbar} J(\omega). \tag{26}$$

The last form is the correct form of the dissipation term in the quantum limit ($k_B T \leq E_R$): This is reminiscent of the Korringa-Shiba relation at low frequency reflecting the Fermi-liquid Kondo ground state [6, 7]:

$$\Im m\chi(\omega) = J(\omega)\Re e\chi(\omega = 0). \tag{27}$$

More precisely, for frequencies $\omega$ in the vicinity of $\omega_R$, we find the following dynamical response:

$$\langle\sigma_z(\omega)\rangle\left(-\omega^2 + \omega_R^2 - i\gamma(\omega)\right) \approx \omega_R \frac{\gamma_l V_l^{in}(\omega)}{\hbar}. \tag{28}$$

The spin dynamical susceptibility for $\omega$ close to $\omega_R$ then takes the underdamped form (see Eq. (14) in the main text):

$$\chi(\omega) \approx \frac{(\omega_R/\hbar)}{\omega_R^2 - \omega^2 - i\gamma(\omega)}. \tag{29}$$

This form of $\chi(\omega)$ is in fact in good agreement with Numerical Renormalization Group results for $0.1 \preceq \alpha \preceq 0.2$ [7] and therefore in the main text we will assume this form of the spin susceptibility for $0.1 \preceq \alpha \preceq 0.2$. Note that this results in the equalities:

$$\Im m\chi(\omega_R) J(\omega_R) = 1 \tag{30}$$
$$\Re e\chi(\omega_R) = 0.$$

Combining these equations with Eqs. (13) in the main text, this shows that the scattering matrix is unitary when $P_{in} \to 0$ for $\omega \sim \omega_R$. Most of the spectrum is elastic close to the resonance.

Eq. (29) is the susceptibility used in the main text, for the underdamped regime ($0.1 \preceq \alpha \preceq 0.2$) and not too low frequencies. Note that in the low-frequency limit $\omega \ll \omega_R$, replacing $\sum_{k>0} -\alpha_k \gamma_l (b_{lk} + b_{lk}^\dagger)\sigma_z/2$ by $-V_l^{in}\gamma_l\sigma_z/2$ would become less accurate.

## NONLINEAR EFFECTS WHEN INCREASING THE INPUT POWER

From Eq. (19), Eq. (29) must be in fact understood as:

$$\chi(\omega) \approx \frac{E_J \langle\sigma_x(t_i)\rangle/\hbar^2}{\omega_R^2 - \omega^2 - i\gamma(\omega)}, \tag{31}$$

To investigate the effect of the input signal amplitude (power), we notice that the result $\langle\sigma_x(t_i)\rangle \to E_R/E_J$ can be easily understood in terms of the polaron transformation resulting in the transformed Hamiltonian $\tilde{H}$ in the main text. The effect of the bath is included only through the modification $\sigma_+ \to \sigma_+ \exp i(\Phi_l - \Phi_r)$ and $\sigma_- \to \sigma_- \exp -i(\Phi_l - \Phi_r)$. Therefore, at sufficiently weak couplings ($\alpha \ll 1/2$) [4] and $T = 0$, this results in $\langle\sigma_x\rangle = \langle\sigma_x\rangle_{\alpha=0} \times \langle\cos(\Phi_l - \Phi_r)\rangle =$ $\langle \cos(\Phi_l - \Phi_r) \rangle$. Again, we have assumed $\epsilon \to 0$. This is the exponential dressing Franck-Condon factor [1]. This implies that Eq. (31) is formally equivalent to:

$$\chi(\omega) = \frac{E_J}{\hbar^2} \frac{\langle \cos(\Phi_l - \Phi_r)(t_i) \rangle}{\omega_R^2 - \omega^2 - i\gamma(\omega)}. \tag{32}$$

We identify $\langle \cos(\Phi_l - \Phi_r) \rangle = \exp -[\langle (\Phi_l - \Phi_r)^2 \rangle/2]$. Now, let us analyze the terms in $\langle (\Phi_l - \Phi_r)^2 \rangle$. One finds:

$$\langle \Phi_l^2 \rangle = -\sum_{k>0} \gamma_l^2 \frac{\alpha_k^2}{(\hbar \omega_k)^2} \langle (b_{lk} - b_{lk}^\dagger)(b_{lk} - b_{lk}^\dagger) \rangle. \tag{33}$$

When $P_{in} \to 0$ and $T \to 0$, we get $\langle b_{lk} b_{lk}^\dagger \rangle = 1$ for all $k$ and similarly for $\langle b_{rk} b_{rk}^\dagger \rangle$ (this is still true for finite $\alpha$ since $\langle \sigma_z \rangle = 0$), and using the adiabatic renormalization [1]:

$$\langle \Phi_l^2 + \Phi_r^2 \rangle = \sum_{k>0} \frac{\lambda_k^2}{(\hbar \omega_k)^2} = \frac{\hbar}{\pi} \int_{\sim E_J/\hbar}^{\omega_c} \frac{J(\omega)}{\hbar^2 \omega^2} d\omega = 2\alpha \ln\left(\frac{\hbar \omega_c}{E_J}\right). \tag{34}$$

Assuming a weak coupling $\alpha$, this allows to recover the result from Bethe Ansatz [4]

$$\langle \cos(\Phi_l - \Phi_r) \rangle \to E_R/E_J. \tag{35}$$

It is relevant to observe that when $P_{in} \to 0$ and $\alpha \ll 1/2$, one can formally replace $\langle \cos(\Phi_l - \Phi_r) \rangle$ by $1 - \langle \Phi_l^2 + \Phi_r^2 \rangle/2$.

Now, let us slightly increase the input signal amplitude; recall, $V_l^{in}(t) = V_0 \cos(\omega t)$ and we are interested in the elastic propagation of a photon with frequency $\omega \sim \omega_R$. The (time-averaged or mean) input power then takes the form $P_{in} = V_0^2/2R = \dot{N}\hbar\omega_R$, where $\dot{N} = dN/dt$ and $N$ represents the number of photons at $x = 0$ in the left transmission line (with an energy $\sim \hbar\omega_R$). Note that following the notations of the main text, one can also identify:

$$P_{in} = \frac{\langle V_l(x=0)^2 \rangle}{R} = \sum_{q>0} \frac{\alpha_q^2}{(2e)^2 R} \langle b_{lq}^\dagger b_{lq} \rangle = \sum_{q>0} \hbar\omega_q (v/\mathcal{L}) \langle b_{lq}^\dagger b_{lq} \rangle. \tag{36}$$

Using the fact that for a monochromatic source

$$\begin{aligned} V_l^{in} &= \frac{V_0}{2} (\exp(i\omega t) + \exp(-i\omega t)) \\ &= \sum_{q>0} \frac{\alpha_q}{2e} (b_{lq}(t) + b_{lq}^\dagger(t)), \end{aligned} \tag{37}$$

and that for the incoming modes $b_{lq}(t) = b_{lq}(0) \exp(-i\omega_q t)$ then, in the sum above, this selects momenta $q$ such that $\omega_q \sim \omega \sim \omega_R$. Therefore, there is a novel contribution to $\langle \Phi_l^2 \rangle$ stemming from $\omega_q \sim \omega_R$ in addition to the high-frequency contribution in Eq. (34). Since the initial time $t_i$ is set arbitrarily (but, at this time the incoming signal has reached the two-level system), when performing different measurements one must also average over $t_i$ and therefore $\langle \Phi_l^2 \rangle$ yields an extra contribution equal to

$$\frac{1}{(\hbar\omega_R)^2} \sum_{q \in '} \alpha_q^2 \gamma_l^2 \langle b_{lq}^\dagger b_{lq} \rangle = \frac{P_{in} R (2e)^2 \gamma_l^2}{(\hbar\omega_R)^2}, \tag{38}$$

and the symbol $'$ refers to momenta such that $\omega_q \sim \omega \sim \omega_R$ in the case of a monochromatic signal. This results in an exponential suppression of $\Im m\chi(\omega = \omega_R)$; see Eqs. (15) and (16) in the main text. Again, this expression assumes that frequencies of interest lie in the range of $\omega_R$. We have also assumed that photons with a (much) larger frequency are still in a thermal equilibrium.

In the underdamped limit, one photon is perfectly transmitted in the time scale $1/\omega_R$ as shown in the main text. We thus define the associated power $P_R = \omega_R(\hbar\omega_R)$. The magnetic susceptibility in the frequency domain $\omega \sim \omega_R$ then reads (assuming that $P_{in} \leq E_J^2/\hbar$; see below)

$$\chi(\omega \sim \omega_R, P_{in}) \approx \frac{(\omega_R/\hbar) \exp -\mathcal{A}}{\omega_R^2 - \omega^2 - i\gamma(\omega)}, \tag{39}$$

where

$$\mathcal{A} = \frac{P_{in}}{P_R} \frac{R}{R_Q} \pi \gamma_l^2. \tag{40}$$

When increasing the input signal amplitude this will produce a macroscopic number of photons with an energy $\hbar\omega_R$: the saturation of the artificial atom excitation manifests itself in a substantial decrease of the photon transmission.

# ASYMPTOTIC FREEDOM OF MICROWAVE LIGHT

In Eqs. (21) and (34), when $P_{in} > E_J^2/\hbar$, note that formally the low-frequency cutoff of the integrals must be changed into $P_{in}/E_J$ ($\hbar P_{in}/E_J$ becomes a large energy scale controlling the departure from $\langle b_{lk} b_{lk}^\dagger \rangle = 1$, *i.e.*, from the ground state) and the characteristic frequency in Eq. (39) becomes

$$\frac{\omega_R(P_{in})}{\hbar} = \frac{E_J}{\hbar}\left(1 - \alpha \ln(E_J \omega_c / P_{in})\right). \tag{41}$$

Finally, we discuss thermal effects. At finite temperatures, performing a thermal average, one finds that $\langle \sigma_x \rangle$ substantially decreases for $k_B T \sim E_R$ [8], since the artificial atom lies in a highly mixed state, producing a strong suppression of the spin susceptibility (this reflects that the reflection coefficient of the microwave light is strongly enhanced). Further, for $(\beta \hbar E_J) \ll 1$, from Eq. (21) we also predict that the characteristic energy of the artificial atom takes the form $E_R(T)/E_J = 1 - \alpha \ln(\beta \hbar \omega_c / 2\pi)$ and $\omega_R$ in the expression (31) then becomes replaced by $E_R(T)/\hbar$. For very high energy scales, we check that $E_R(T)$ would converge to the bare value $E_J$ in the Hamiltonian $\tilde{H}$.

These two facts exemplify the asymptotic freedom where microwave light and two-level system almost disentangle. Note the parallel between temperature and driving effects in the spin susceptibility close to the confinement frequency.

---